# Electrically Enhanced Exchange Bias via Solid State Magneto-Ionics

Peyton D. Murray[1], Christopher J. Jensen,[2] Alberto Quintana,[2] Junwei Zhang[3], Xixiang Zhang[3],

Alexander J. Grutter,[4] Brian J. Kirby,[4] and Kai Liu[1,2*]

[1]*Physics Department, University of California, Davis, CA 95616, USA*
[2]*Physics Department, Georgetown University, Washington, DC 20057, USA*
[3]*King Abdullah University of Science & Technology, Thuwal 23955-6900, Saudi Arabia*
[4]*NIST Center for Neutron Research, National Institute of Standards and Technology, Gaithersburg, MD 20899, USA*

## Abstract

Electrically induced ionic motion offers a new way to realize voltage-controlled magnetism, opening the door to a new generation of logic, sensor, and data storage technologies. Here, we demonstrate an effective approach to magneto-ionically and electrically tune exchange bias in Gd/Ni$_{1-x}$Co$_x$O thin films ($x$=0.50, 0.67), where neither of the layers alone is ferromagnetic at room temperature. The Gd capping layer deposited onto antiferromagnetic Ni$_{1-x}$Co$_x$O initiates a solid-state redox reaction that reduces an interfacial region of the oxide to ferromagnetic NiCo. Exchange bias is established after field cooling, which can be enhanced by up to 35% after a voltage conditioning and subsequently reset with a second field cooling. These effects are caused by the presence of an interfacial ferromagnetic NiCo layer, which further alloys with the Gd layer upon field cooling and voltage application, as confirmed by electron microscopy and polarized neutron reflectometry studies. These results highlight the viability of the solid-state magneto-ionic approach to achieve electric control of exchange bias, with potentials for energy-efficient magneto-ionic devices.

*Corresponding author: Kai.Liu@georgetown.edu

## INTRODUCTION

Methods aiming at controlling magnetism using electric fields have attracted significant interests in recent years with the promise of enabling a new generation of nonvolatile, low-dissipation electronics,[1-9] bypassing the joule heating effect associated with electric currents in conventional systems. In this regard, atomic scale control of interfaces via ionic migration in solid state heterostructures[6-13] as well as electrolyte-based systems[2, 14-20] has emerged as an effective tool to modify materials properties, which can be further controlled by an electric field. To date, a variety of magneto-ionically controlled functionalities have been demonstrated, including magnetic anisotropy,[7-8, 13] antiferromagnetism,[12] ferromagnetism,[9, 14, 18, 20-23] superconductivity,[24-25] Dzyaloshinskii-Moriya interaction (DMI) and spin textures,[26-28] etc. Much of the magneto-ionic studies have focused on the interfacial electrostatic effect,[2, 6-8] where the charge build-up across interfaces under electric field modifies electronic structures and materials characteristics. Recent studies have also demonstrated that the electrochemical means, where the interface is transformed by an ion-migration-induced chemical reaction, can modify materials properties well beyond the interface.[12, 15, 21, 24, 29] For example, redox reaction at a ferrimagnetic-GdFe / antiferromagnetic-NiCoO interface has been shown to trigger oxygen ion migration, which is manifested in the exchange bias effect.[12] Conversely, this magneto-ionic effect offers a new paradigm to explore the potential for electrical manipulation of exchange bias. As exchange bias is central to spin-valve type devices such as magnetic tunnel junctions that enable control of magnetic configurations,[30-32] electric control of exchange bias may lead to substantially more energy-efficient switching than using a magnetic field. Indeed, there has been keen interest in exploring electric switching of exchange bias, using multiferroics,[33-34] memristors,[35] or electrolytes.[16]

In this work, we present an effective approach to magneto-ionically and electrically tune exchange bias in Gd/Ni$_{1-x}$Co$_x$O thin films ($x$=0.50, 0.67). Even though neither of the constituents is ferromagnetic at room temperature, the Gd layer strips oxygen away from the antiferromagnetic (AF) Ni$_{1-x}$Co$_x$O (referred to as NiCoO), reducing its interface region to a ferromagnetic (FM) layer. A significant exchange bias is established upon field cooling, which can be further enhanced in an electric field, and reset by a subsequent field cooling. Electron microscopy and polarized neutron reflectometry (PNR) studies provide direct evidence of the formation of a FM NiCo interfacial layer and reveal microstructural modifications of the interface upon field-cooling and electric field gating. Our approach combines the strong ionic migration induced by the Gd layer with electrical biasing to allow for on-demand modification of exchange bias, opening up an effective magneto-ionic pathway towards nonvolatile and energy efficient magnetic switching devices.

**SAMPLE SYNTHESIS**

Polycrystalline thin film samples of Gd/Ni$_{0.50}$Co$_{0.50}$O (series 1) and Gd/Ni$_{0.33}$Co$_{0.67}$O (series 2) were magnetron sputtered in an ultrahigh vacuum system with a base pressure in the $10^{-6}$ Pa range. The Gd is a strong oxygen getter material,[21, 29] can extract oxygen from deep within an adjacent oxide layer due to its low electron work function,[24] with a Curie temperature of 292K.[36] The AF NiCoO has a tunable Néel temperature that scales with the composition:[37] T$_N$ = 408 K for Ni$_{0.50}$Co$_{0.50}$O and 368K for Ni$_{0.33}$Co$_{0.67}$O.[12, 38] For series 1, a naturally oxidized Si wafer was first sputter-coated with a Ta (1 nm thick) adhesion layer followed by a Pt (10 nm) layer, forming a bottom electrode. The wafer was then removed from the deposition chamber and lithographically patterned into arrays of 7.5 mm × 7.5 mm squares uncovered by the resist. Subsequently, an insulating layer of Al$_2$O$_3$ (170 nm) was radio frequency (RF) sputtered from an Al$_2$O$_3$ target with a working gas consisting of 2% O$_2$ / 98% Ar. Next, a layer of Ni$_{0.50}$Co$_{0.50}$O (20 nm) was RF-sputtered from a stoichiometrically balanced target in 0.33 Pa of a 15% $O_2$ / 85% Ar working gas mixture. Finally, the sample was capped with a Gd layer (20 nm) followed by a layer of Pt (10 nm), with the latter acting both as a top electrode and as a barrier to protect the sample from atmospheric exposure. Except for the NiCoO layer, sputtering was carried out at a gas pressure of 0.67 Pa. After deposition, the sample was washed in acetone to remove the remaining photoresist, exposing the bottom electrode. The final sample structure is illustrated in Fig. 1.

Series 2 samples were deposited by reactive magnetron sputtering onto thermally oxidized $SiO_2$ (285 nm) grown on $p$-type Si, which acts as a bottom electrode. The Ar : $O_2$ fraction was optimized following reported procedures.[39-40] A $Ni_{0.33}Co_{0.67}O$ (40 nm) layer was then reactively co-sputtered using elemental Ni and Co targets in a 0.33 Pa gas mixture of 6.7% $O_2$ / 93.3% Ar, at a substrate temperature of 500 °C.[41] Patterned 5mm×5mm samples were obtained using a mask during the sputtering process. Subsequently, a 20 nm layer of Gd and a 20 nm Pd capping layer were deposited onto the films. This second series was chosen as exchange bias in $Ni_{1-x}Co_xO$ based systems is composition sensitive, due to the competition between the higher $T_N$ of NiO and the larger anisotropy of CoO.[42]

**MAGNETOMETRY**

Vibrating sample magnetometry (VSM) measurements of series 1 samples were first carried out in the as-grown (AG) state. A nonzero magnetization is observed, along with a coercivity of 168 Oe (Fig. 2a), even though none of the constituent materials alone exhibits ferromagnetism at room temperature. This indicates that a partial reduction of $Ni_{0.50}Co_{0.50}O$ has occurred, due to the gettering effect of the Gd layer that leads to a ferromagnetic layer at the interface, similar to that observed previously in the GdFe/NiCoO system.[12] The equivalent thickness, $t$, of a continuous ferromagnetic layer corresponding to the observed moment can be estimated from the saturation magnetization of Ni (495 emu / cm³) [1 emu / cm³= 1 kA m⁻¹] and Co (1400 emu / cm³), depending on the exact Ni:Co ratio; for these samples, 0.4 nm $\leq t \leq$ 1.3 nm, with the lower and upper bounds determined assuming pure Co and pure Ni, respectively. More details about this layer will be discussed below in the neutron studies.

A second measurement was made after heating the samples in Ar above the $Ni_{0.50}Co_{0.50}O$ $T_N$ to 420 K, and then cooling to room temperature in a 10 kOe [1 kOe = 0.1 T / μₒ] magnetic field. After this field cooling (FC) step, an exchange bias is established with a bias field $H_b$ = –283 Oe and an enhanced coercivity of 221 Oe (Fig. 2a).

A third measurement was made after a voltage conditioning procedure, wherein a 10 V bias (~0.5 MV cm⁻¹ electric field) was applied along the $-\hat{z}$ direction (Fig. 1) for 12 hours at room temperature, with the $O^{2-}$ ions in the $Ni_{0.50}Co_{0.50}O$ layer expected to drift towards Gd. Magnetometry measurements of this voltage-conditioned (FC+VC) state show an increase in the exchange bias to $H_b$ = – 308 Oe, along with a further coercivity enhancement to $H_C$ = 230 Oe (Fig. 2a). Subsequently, a reverse-bias voltage was applied at room temperature using the same 0.5 MV/cm electric field applied for 12 hours in the $+\hat{z}$ direction. This reverse-bias voltage conditioned (FC+VC-VC) state showed no appreciable change in the hysteresis loop.

Finally, in a second field cooling procedure, the sample was again heated to 420 K and then cooled to room temperature in a 10 kOe field. A subsequent magnetometry measurement (FC+VC+FC) shows the major loop returning from the FC+VC state back to the original FC state (Fig. 2a), and the electric field induced exchange bias is reset.

Magnetometry measurements for series 2 samples were carried out in a superconducting quantum interferometer device magnetometer, following similar procedures as those for series 1. The AG sample again exhibited a ferromagnet like hysteresis loop, similar to series 1, with a coercivity of $H_C = 260$ Oe (Fig. 2b). Subsequently, a field cooling process was carried out by increasing the temperature to 400 K, above the $Ni_{0.33}Co_{0.67}O$ $T_N$ of 368 K, and cooling down to room temperature in an external field of 10 kOe. An exchange bias of $H_b = -80$ Oe was established, along with an enhanced coercivity of $H_C = 276$ Oe. This FC sample was then subjected to a voltage gating of 20V ($\sim 0.6$ MV/cm) for 12h. The exchange bias was enhanced to $H_b = -108$ Oe, or a 35% increase upon voltage treatment, and the coercivity further increased 297 Oe. As in series 1 samples, reverse biasing produced no appreciable effect on the sample, while a second field-cooling procedure can reset the magnetic state of the sample.

## ABERRATION-CORRECTED SCANNING TRANSMISSION ELECTRON MICROSCOPY

To further understand the sample microstructure under different conditions, aberration-corrected scanning transmission electron microscopy (Cs-corrected STEM) images were acquired at KAUST on three $Gd/Ni_{0.50}Co_{0.50}O$ samples prepared in the AG, FC, and FC+VC states. Representative images of the $Gd/Ni_{0.50}Co_{0.50}O$ interface show small, randomly oriented crystallites identifiable as regions of uniform and parallel lattice fringes, indicating that both Gd and $Ni_{0.50}Co_{0.50}O$ form polycrystalline grains during growth (Fig. 3a). Some $NiCo_2O_4$ crystallites are also observed, as a result of oxygen leaching from the $Ni_{0.50}Co_{0.50}O$. After field cooling, only a few crystal planes are visible; as thermal treatment often improves crystallinity,[20] this increased disorder may be due to disrupted interfaces caused by ionic migration, which is enhanced by the elevated temperatures reached during field cooling (Fig. 3b). Notably, some crystallites of Gd-Ni alloys are observed at the interface region. After voltage conditioning the imaged region remains largely disordered, and the interface becomes even less sharp (Fig. 3c). While $Ni_{0.50}Co_{0.50}O$ grains are visible in the AG and FC samples, none is visible in the FC+VC sample. Since ionic diffusion is expected to be enhanced at grain boundaries,[43-45] the disruption of the $Ni_{0.50}Co_{0.50}O$ grains is likely responsible for the irreversibility of the voltage-conditioning effects under reverse bias.

Elemental distributions of Co (cyan), Ni (pink), Gd (dark-blue), and O (green) across the $Gd/Ni_{0.50}Co_{0.50}O$ interface were further obtained by using electron energy loss spectroscopy (EELS) elemental mapping for the AG, FC and FC+VC state, as shown in Fig. 4, with the Gd layer shown at the left end, and $Ni_{0.50}Co_{0.50}O$ at the right end. In the AG sample, a weak oxygen signal is observed in the Gd region, suggesting partial oxidation of the Gd. A clear Ni and Co region, free of oxygen, is observed at the $Gd/Ni_{0.50}Co_{0.50}O$ interface (Fig. 4a), in agreement with magnetometry results. These observations therefore constitute a direct confirmation of the presence of the NiCo metallic layer proposed previously,[12] which is expected to form as a result of Gd-induced oxygen leaching. The oxygen signal is recovered more in the Ni and Co region (right side), as expected for the $Ni_{0.50}Co_{0.50}O$ layer. In the FC state, the oxygen concentration remains low at the interface. A clear broadening of the oxygen-free Ni and Co interface is observed, along with a partial overlap between Gd and Ni. The latter chemical segregation arises as a result of the lower alloying enthalpy of formation of Gd-Ni alloys as compared to Gd-Co alloys.[46-47] During the initial field cooling, oxygen migration from the $Ni_{0.50}Co_{0.50}O$ to the Gd layer leaves behind metallic NiCo; as a result of Gd-Ni alloys being energetically favored over Gd-Co, the Ni atoms are drawn deeper into the Gd layer and away from the remaining NiCoO layer, accounting for the Ni enrichment near the interface. Additionally, oxygen concentration decreases in the Gd layer near the interface, suggesting that oxygen has penetrated deeper into the bulk of Gd (Fig. 4b). Upon voltage conditioning further segregation between the Ni and Co elemental distributions is observed, beyond that observed after FC (Fig. 4c). In this case, however, the interfacial layer right next to Gd is almost entirely composed of Ni, again pointing to the formation of a Gd-Ni alloy and suggesting that the disorder induced by ionic migration under voltage conditioning acts to further enhance the chemical segregation effect. Once again, the oxygen is absent at the interface and its presence is only significant deep into the NiCoO layer.

**POLARIZED NEUTRON REFLECTOMETRY**

Structural and magnetic depth profiles of series 2 samples have been further analyzed by PNR over macroscopic areas,[48-49] as shown in Figs. 5a, c, e in terms of the real ($\rho_N$) and imaginary component ($\rho_i$) of the nuclear scattering length density (SLD) and the magnetic scattering length density ($\rho_M$).[7] The Gd layer, a strong neutron absorber, is easy to identify during modelling due to its large $\rho_i$.[7] In addition, $O^{2-}$ migration into Gd can be tracked with increases in $\rho_N$ and decreases in $\rho_i$ with the formation of GdO$_x$.[11] For all the measured samples (AG, FC, and FC+VC) it was determined that a FM interfacial layer must be present, in agreement with magnetometry results discussed above (Fig. 2), as highlighted by the fitted reflectivity (Supporting Information, Figure S1) and spin asymmetry results (Figs. 5b, d, f).

In the AG sample, PNR reveals an interfacial NiCo layer between the NiCoO and Gd (Supporting Information, Figure S2). The thickness of the NiCo layer is modeled to be 3.2 nm and exhibits a $\rho_N$ that matches the expected value of $4.7 \times 10^{-4} \, nm^{-2}$, calculated from a 2:1 composition of Co : Ni at the interface (Fig. 5a). The measured magnetic scattering length density, $\rho_M$, of $0.45 \times 10^{-4} \, nm^{-2}$ is smaller than the theoretical value of $3.17 \times 10^{-4} \, nm^{-2}$. The apparently slightly thicker NiCo layer than that observed in STEM and expected from magnetometry, along with the smaller value of $\rho_M$, can be attributed to the presence of a discontinuous, non-uniform interfacial NiCo layer with non-FM regions. Adjacent to the NiCo interface is a GdO$_x$ layer with a negligible $\rho_M$, which can be distinguished from the Gd layer in the $\rho_N$ and $\rho_i$ values. In the Gd layer, $\rho_N$ and $\rho_i$ were best modeled using a continuous value throughout the layer thickness, with increased interface roughness between Gd and Pd. A non-zero $\rho_M$ is observed, and is attributed to the measurement temperature being near Gd $T_C$ and strain effects (Supporting Information).

For the FC sample, PNR reveals a thinner 2.9 nm NiCo layer at the NiCoO/Gd interface, with a higher $\rho_M$ of $0.75 \times 10^{-4} \ nm^{-2}$ (Fig. 5c), suggesting a more compact NiCo layer compared to that in the AG sample. Additionally, there is a second interfacial FM layer, differentiable from the $GdO_x$ layer in the AG sample by an increase in $\rho_N$, dramatic decrease in $\rho_i$, and inclusion of a $\rho_M$, corresponding to Ni migration and formation of a Gd-Ni alloy of 4.5 nm. Over much of the Gd layer, the homogeneous $\rho_i$ and $\rho_N$ distribution suggest that the FC process may have promoted a redistribution of oxygen along the entire Gd film, evidenced by an increase in $\rho_N$ from $0.80 \times 10^{-4} \ nm^{-2}$ to $1.35 \times 10^{-4} \ nm^{-2}$. For the FC+VC sample, the interfacial NiCo layer thickness is further reduced to 2.1 nm, with a $\rho_M$ of $0.94 \times 10^{-4} \ nm^{-2}$. The second interface of Gd-Ni alloy increases in thickness up to 4.7 nm and $\rho_N$ also increases, indicating further Ni alloying with Gd after voltage conditioning (Fig. 5e). In the Gd layer, a further increase in oxygen content is apparent with an increase in $\rho_N$ to $1.83 \times 10^{-4} \ nm^{-2}$, consistent with direction of $O^{2-}$ migration under voltage application. Note that depth profiles with a single FM layer of NiCo, without the GdNi layer, were also studied, which did not show as good fits (Supporting Information Figures S3 and S4).

## DISCUSSIONS

The observed magnetic behavior of the Gd/$Ni_{1-x}Co_xO$ system is the result of two separate ion migration mechanisms. The first occurs upon the Gd layer deposition, resulting in a chemically-induced redox reaction that strips oxygen from the adjacent NiCoO and leaves behind the FM NiCo observed in magnetometry, PNR and STEM of the as-grown samples. The resultant valence change at the interface is similar to those reported earlier.[21, 24] More disorder is induced at the NiCoO top surface after the Gd deposition (Supporting Information Figure S5). This gadolinium oxidation process is exothermic, with a change in enthalpy $\Delta H$ of –11.4 eV per molecule of $Gd_2O_3$ formed $[3(Ni, Co)O + 2Gd \rightarrow 3(Ni, Co) + Gd_2O_3]$.[50] Thus, local heating may help to increase ionic mobility in the surrounding region, although the degree to which this feedback effect enhances the redox reaction is difficult to quantify due to the inhomogeneity of the interface. The magnetization of this structurally disordered interfacial NiCo region is lower than expected for a nominal NiCo alloy, as confirmed by the lower $\rho_M$ value extracted from PNR (Fig. 3a), owing to residual oxygen content and different atomic coordination as compared to the crystalline phase. After the FC process, the enhanced ionic mobility associated with elevated temperatures combined with the Gd-driven chemical potential gradient induces additional oxygen migration beyond that of the as-grown state, and further broadens the interfacial region, as observed in the STEM image.

The second ion migration mechanism is the electric field induced motion. During voltage conditioning, oxygen ions migrate toward the Gd layer, acting to mix and further disorder the interfacial region. Consequently, the NiCo alloy becomes increasingly Ni-enriched via the alloying-enthalpy mechanism discussed earlier. As the relative ratio of Ni ($M_s = 495$ emu/cm$^3$) to Co ($M_s = 1400$ emu/cm$^3$)[36] increases, the magnetization of the interfacial layer decreases below that of the FC state. Additionally, the thickness of this layer may also change as a result of the introduction of structural disorder. A number of microscopic accounts of the exchange bias, such as the random field model,[51-52] show that the observed exchange bias is inversely proportional to both the saturation magnetization and the thickness of the ferromagnetic layer:

$$H_E = \frac{2z\sqrt{AK}}{\pi^2 M_s t}$$

where $A$ and $K$ are the AF layer exchange stiffness and anisotropy, respectively; $M_s$ and $t$ are the FM saturation magnetization and thickness; and $z$ is a number of order unity which depends on the shape of the AF domains. The increase in the bias observed in the hysteresis loops of the FC and FC+VC states suggest that the quantity $M_s t$,[53] which corresponds to the saturation magnetization and thickness of the FM NiCo at the interface, decreases as a result of voltage conditioning. This may be due to a decrease in $M_s$ from Ni enrichment, or a decrease in the *effective* FM layer thickness, as oxygen migration through the interfacial FM layer alters the structure. This oxygen migration away from the NiCoO likely also leads to an increase in the density of pinned uncompensated AF spins at the FM/AF interface, which directly relates to exchange bias.[54] These effects act to increase the exchange bias field, in agreement with the measured hysteresis loops of the FC and FC+VC states for both sets of samples.

One possible explanation for the reversibility of the field-induced exchange bias enhancement through FC but not inverse bias points to the relative ionic diffusivities along grain boundaries and through the bulk. At room temperature, diffusion along grain boundaries is typically energetically favored over other pathways;[43] at elevated temperatures, bulk diffusion becomes important, and both the bulk and grain boundary diffusivities converge. Upon field cooling of a pristine sample, the elevated temperatures reached during the thermal cycling is therefore expected to enhance both diffusivities, allowing oxygen migration from the $Ni_{1-x}Co_xO$ toward the Gd from the bulk as well as along grain boundaries. The ionic migrations act to disorder the $Ni_{1-x}Co_xO$ near its interface with Gd, as can be seen in Fig. 4b, but does not fully disrupt the $Ni_{1-x}Co_xO$/Gd grain boundaries, which are still visible in the image. Subsequent room temperature voltage conditioning, however, is expected to induce ionic migration primarily along grain boundaries; thus, most disruptions and disordering due to ionic migration are expected at the grain boundaries. Indeed, in the FC+VC STEM image (Fig. 3c), only a few crystalline planes are observed over nanometer scale regions whose boundaries are highly disordered. Since ion mobility is expected to be highest along these grain boundary pathways, disruption to these pathways renders the ionic distributions immobile under reverse bias, meaning that voltage-biasing in this material may be expected to only cause significant ionic migration under the first voltage-conditioning treatment. Importantly, a second field cooling procedure can reset the exchange bias back to its original value. Since bulk ionic diffusivity is enhanced at elevated temperature, grain boundaries are not as important for ionic migration during field cooling, allowing the interface to recover to the pre-voltage-conditioned state.

**CONCLUSIONS**

In summary, effective magneto-ionic control and electric field enhancement of exchange bias in $Ni_{1-x}Co_xO/Gd$ have been demonstrated. Ferromagnetism emerges in this system, featuring constituents that are not ferromagnetic at room temperature, due to the formation of a NiCo layer at the $Ni_{1-x}Co_xO/Gd$ interface. This ferromagnetic layer is observed by magnetometry and PNR, and directly confirmed by EELS. This is the result of oxygen ion migration induced by a Gd-oxidation reaction. Exchange bias is established via a field cooling process, which can be further enhanced up to 35% under an electric field gating. EELS and PNR results confirm that upon field cooling, the interfacial NiCo region becomes broader and Ni-rich. Upon voltage conditioning, the reduction of the interfacial NiCo moment due to the additional formation of GdNi alloy, and disorder-induced changes to the effective thickness of the ferromagnetic layer and density of pinned uncompensated AF spins, contribute to the enhancement of the exchange bias. Finally, this effect was found to be reversible under thermal cycling but not under reverse biasing, due to ionic-migration-induced disruption of grain boundaries near the interface, which comprise the most effective pathways for ionic motion. These results demonstrate a new magneto-ionic approach towards electric manipulation of exchange bias, which is highly relevant for energy-efficient spintronic devices.

**Supporting Information**.

Methods, polarized neutron reflectometry measurements of the $Gd/Ni_{0.33}Co_{0.67}O$ samples in AG, FC, and FC+VC states, and x-ray reflectivity measurements (PDF).

**Acknowledgements.** This work has been supported in part by the NSF (ECCS-1611424 and ECCS-1933527), by SMART, one of seven centers of nCORE, a Semiconductor Research Corporation program, sponsored by National Institute of Standards and Technology (NIST), and by KAUST (OSR-2019-CRG8-4081). The acquisition of a Magnetic Property Measurements System (MPMS3) at GU which was used in this investigation was supported by the NSF (DMR-1828420). We thank Professor Yayoi Takamura for helpful discussions.

**Figure Captions**

**Fig. 1.** Layer structure of the samples, with the height of the patterned $Al_2O_3$/NiCoO/Gd/Pt structures exaggerated. Inset shows a magnified view of the NiCoO/Gd interface, with the emergent metallic NiCo layer explicitly shown. The arrow denotes the $+\hat{z}$ direction.

**Fig. 2.** (a) Hysteresis loops of a Gd/$Ni_{0.50}Co_{0.50}O$ sample in the as-grown (AG) state, after field cooling (FC), after voltage conditioning (FC+VC), and after a second field cooling (FC+VC+FC). (b) Hysteresis loops of a Gd/$Ni_{0.33}Co_{0.67}O$ sample in the as-grown (AG) state, after field cooling (FC), after voltage conditioning (FC+VC). Insets show zoomed-in views to highlight changes in exchange bias.

**Fig. 3.** STEM images of the Gd/$Ni_{0.50}Co_{0.50}O$ sample in the (a) as-grown state, (b) after field cooling, and (c) after field cooling + voltage conditioning, with Gd and NiCoO on the left and right side of the interface, respectively. Yellow dashed lines highlight the interface region, including the presence of crystalline GdNi phase in (b) and (c).

**Fig. 4.** EELS study of the Gd/$Ni_{0.50}Co_{0.50}O$ sample, showing elemental distribution maps of Gd (dark blue), Ni (pink), Co (cyan), and O (green) for the (a) as-grown state, (b) after field cooling, and (c) after field cooling + voltage conditioning, with the corresponding high-angle annular dark-field (HAADF) image stacked on top. Yellow dashed lines highlight the elemental segregation in the interface region between Gd and NiCoO, showing a CoNi region in (a), and GdNi and CoNi regions in (b) and (c).

**Fig. 5.** Nuclear and magnetic depth profiles of the Gd/Ni$_{0.33}$Co$_{0.67}$O sample in (a) as grown, (c) field cooled, and (e) voltage conditioned state, showing the real ($\rho_N$), imaginary ($\rho_i$), and magnetic ($\rho_M$) component of the nuclear scattering length density with a black solid curve, black dashed curve, and blue solid curve, respectively. Colored regions are used to indicated various modelled layers: SiO$_2$ (grey), Ni$_{0.33}$Co$_{0.67}$O (teal), NiCo (light green), GdO$_x$ (dark orange), GdNi (dark green), Gd (light orange), Pd (magenta), and air (white). The corresponding measured spin asymmetries and the fits for the accepted models are shown in (b), (d), and (f), respectively.

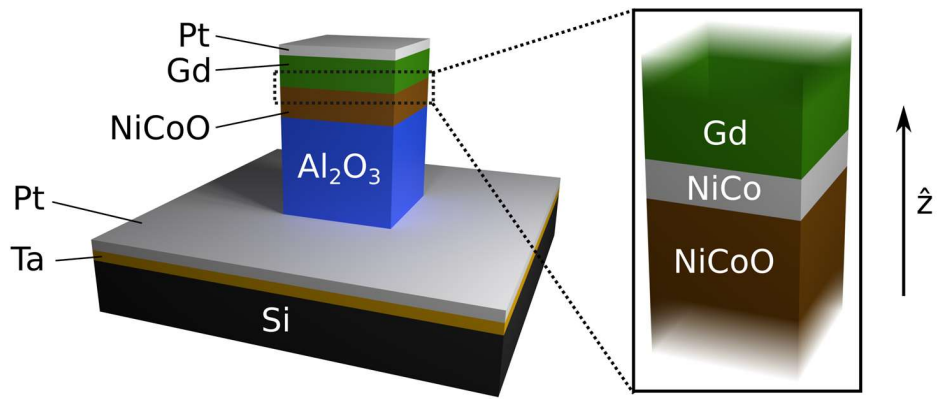

**Fig. 1**

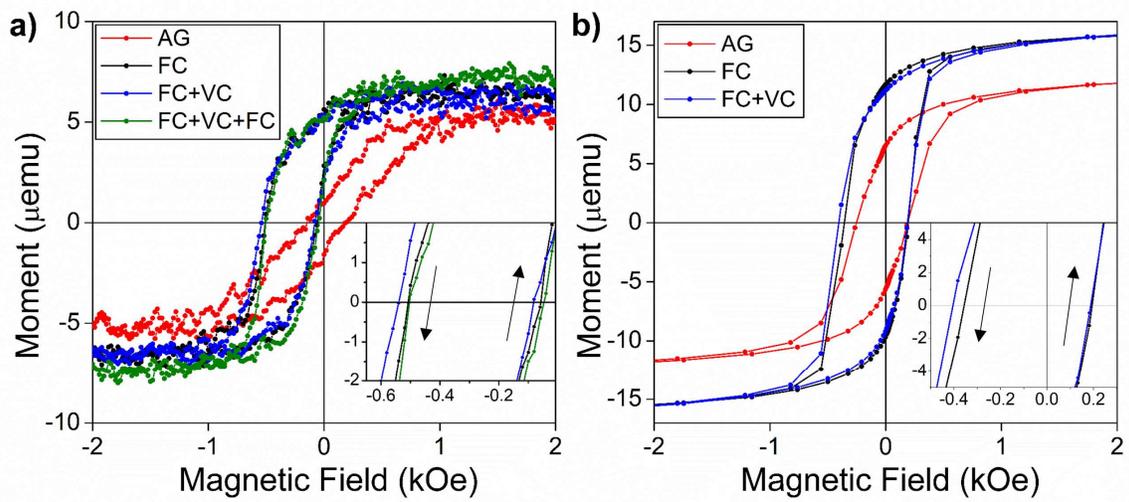

**Fig. 2**

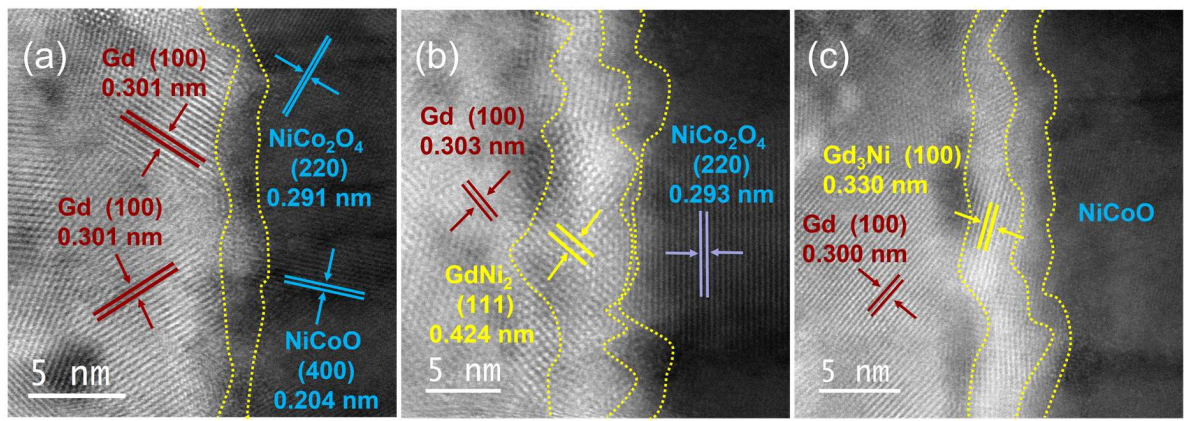

Fig. 3

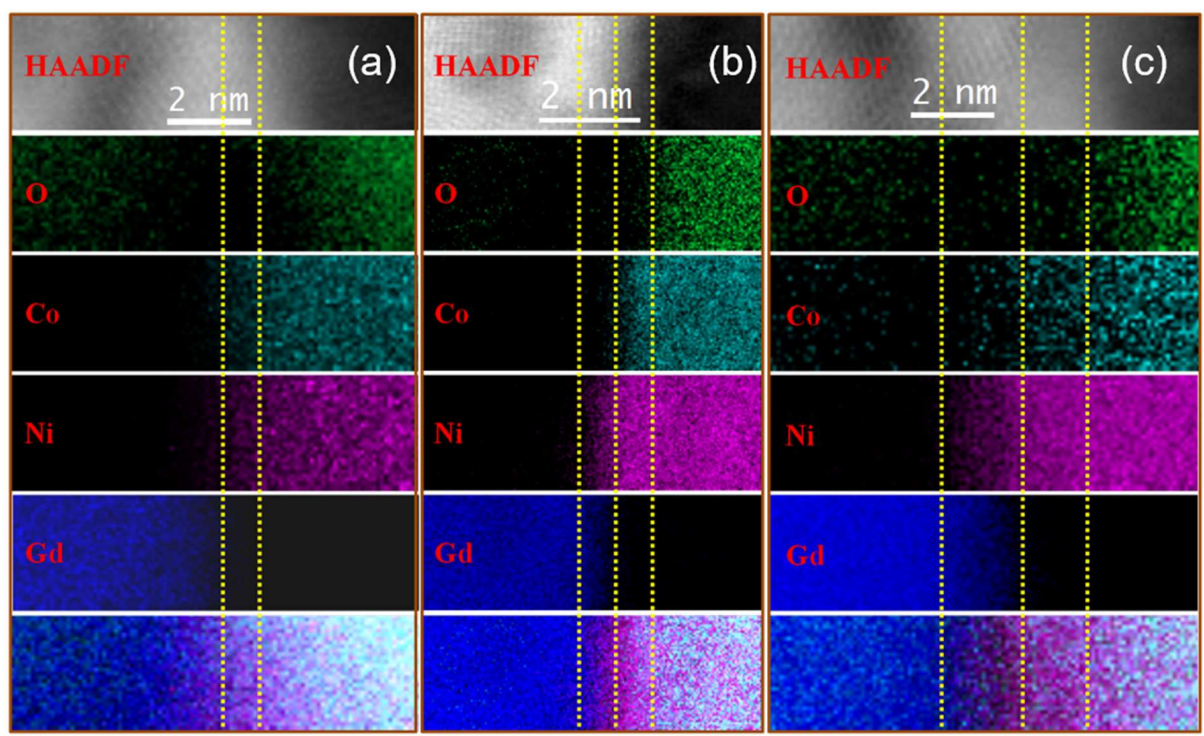

Fig. 4

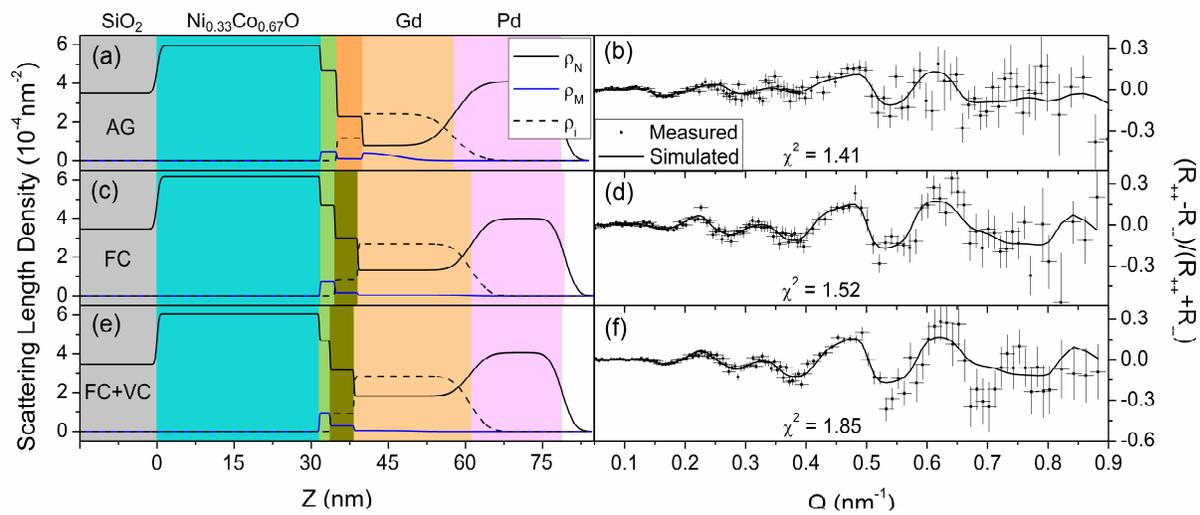

Fig. 5

## Electrically Enhanced Exchange Bias via Solid State Magneto-Ionics

Peyton D. Murray[1], Christopher J. Jensen,[2] Alberto Quintana,[2] Junwei Zhang[3], Xixiang Zhang[3],

Alexander J. Grutter,[4] Brian J. Kirby,[4] and Kai Liu[1,2*]

[1]*Physics Department, University of California, Davis, CA 95616, USA*
[2]*Physics Department, Georgetown University, Washington, DC 20057, USA*
[3]*King Abdullah University of Science & Technology, Thuwal 23955-6900, Saudi Arabia*
[4]*NIST Center for Neutron Research, National Institute of Standards and Technology,
Gaithersburg, MD 20899, USA*

*Corresponding author: Kai.Liu@georgetown.edu

## Methods

## <u>Sample characterization</u>

PNR measurements were performed at the NIST Center for Neutron Research (NCNR) on the PBR reflectometer at room temperature with an in plane applied magnetic field of 16 kOe [1 kOe = 0.1 T / $\mu_o$]. Incident (scattered) neutrons were polarized with their spin parallel (+) or antiparallel (-) to the field, and the non-spin-flip specular reflectivities ($R_{++}$ and $R_{--}$) were measured as a function of wave vector transfer. Spin asymmetry, SA = ($R_{++}$ - $R_{--}$)/($R_{++}$ + $R_{--}$) is a useful representation for visualizing magnetic contributions to the reflectivities. Additionally, X-ray reflectivity (XRR) studies were performed on a Malvern-Panalytical X'Pert3 MRD system with Cu K$_\alpha$ radiation.

Magnetometry studies were carried out using a Princeton Measurement Corporation Vibrating Sample Magnetometer (VSM) MicroMag 3900 and a Quantum Design Superconducting Quantum Interferometer Device (SQUID) magnetometer MPMS3.[1]

## Polarized Neutron Reflectometry

## Supporting Information

Gd/Ni$_{0.33}$Co$_{0.67}$O samples were prepared in the as grown (AG), field cooled (FC), and voltage conditioned state (0.6 MV/cm for 12h, FC+VC). Afterwards, samples were saturated at room temperature in a 1.6 T magnetic field which was directed along the field cooled axis. Incident neutrons were polarized in the parallel and antiparallel direction to H. Measurements of the non-spin-flip cross sections (R$_{++}$ and R$_{--}$) were taken with respect to momentum transfer (Q) normal to the sample surface (Figure S1). Data was reduced and modeled using the REDUCTUS and REFL1D software packages.[2-3]

### Best fits to the PNR data

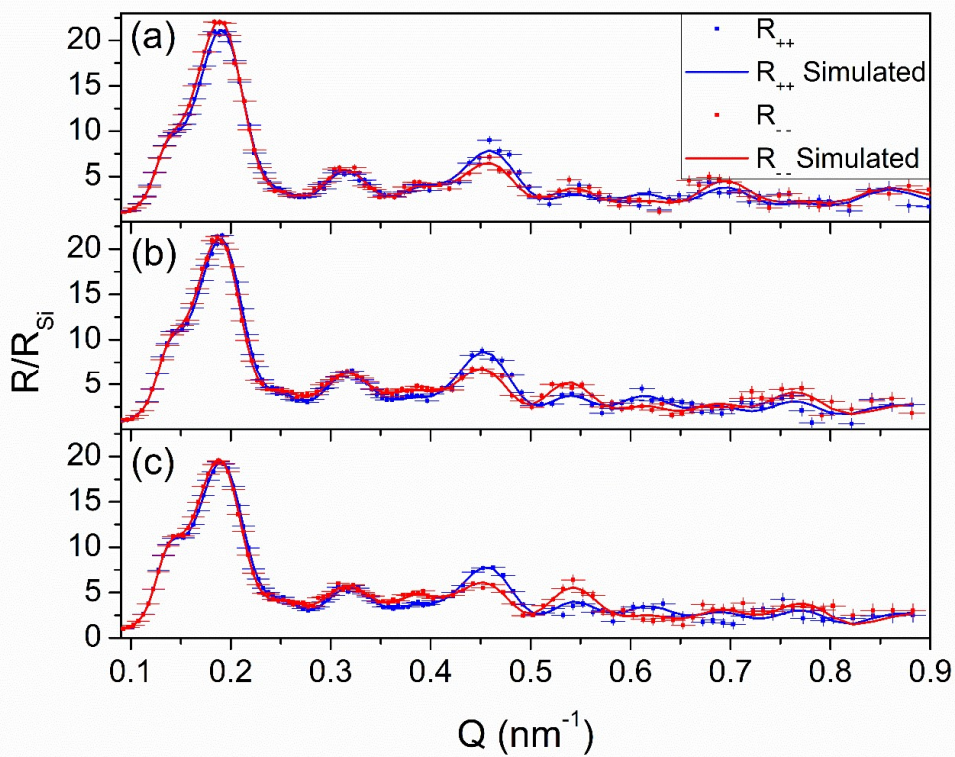

**Figure S1**: Fresnel normalized data and optimal fittings, in (a) as grown, (b) field-cooled, and (c) voltage conditioned state for the series 2 (Ni$_{0.33}$Co$_{0.67}$O) samples. Error bars correspond to ±1 s.d.

### PNR Fitting Considerations and Model Development

PNR is a powerful nondestructive technique that allows evaluation of both structural and magnetic changes along the sample thickness. Real space information is inferred from model fitting of the PNR data. As such, care must be taken to ensure that the data actually provides sensitivity to parameters of interest. For this reason, it is important to start with the simplest case and evolve it towards a more complex model that makes physical sense. In the present system it is important to use this sequence of fits to justify inclusions like the ferromagnetic (FM) interfacial layers, track oxygen distribution throughout Gd, and to avoid inclusion of layers in our model that the data do not actually provide sensitivity to. In the following section, some of the tested models used to converge on a solution and proper model determination will be described.

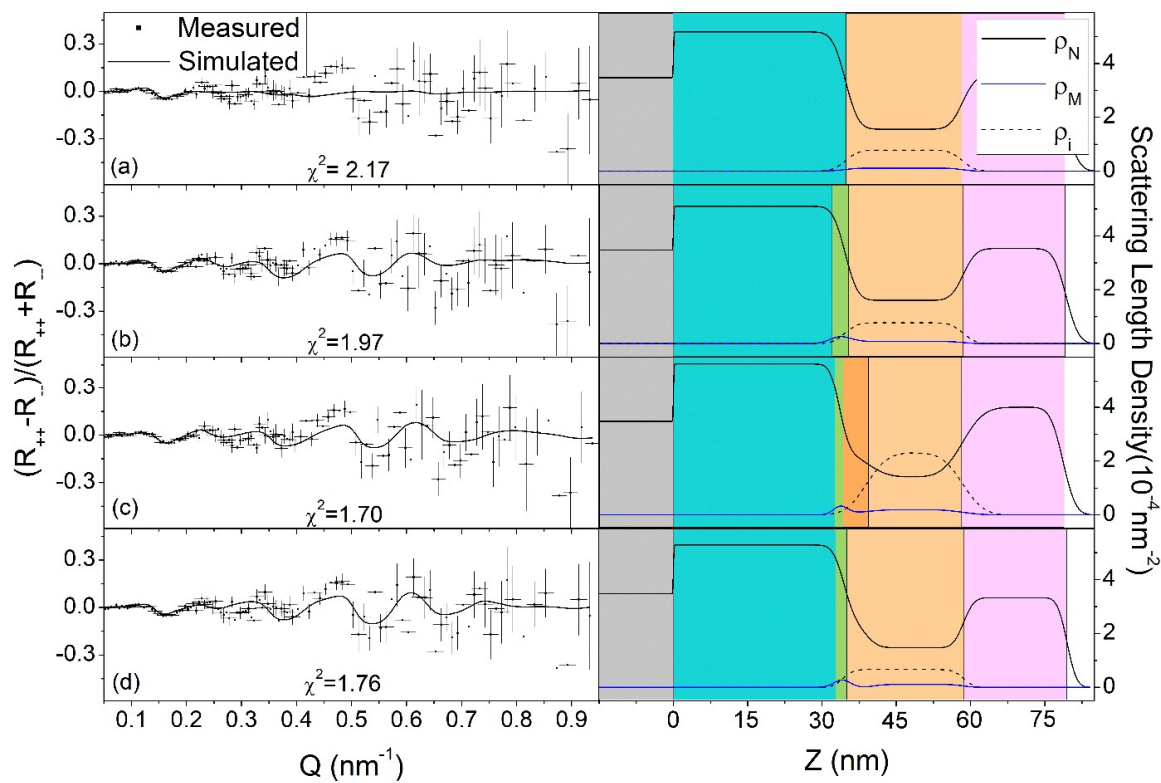

**Figure S2:** Spin asymmetry (left column) and calculated profiles (right column) for different fitting configurations for the $Ni_{0.33}Co_{0.67}O/Gd/Pd$ system in the as grown state. (a) Model not including a FM NiCo interfacial layer, (b) with the inclusion of a FM NiCo interfacial layer, (c) inclusion of a $GdO_x$ layer at the interface adjacent to the NiCo, and (d) a linear gradient in composition for $GdO_x$. Colored regions correspond to (left to right): $SiO_2$ (grey), $Ni_{1-x}Co_xO$ (teal), NiCo (green), Gd (light orange), $GdO_x$ (dark orange), Pd (magenta) and air (white). Error bars correspond to ±1 s.d.

**Supporting Information**

Figure S2 shows the spin asymmetry (left column) and calculated profiles (right column) of 4 models of the as grown, AG, sample. In Figure S2a, the data was fit without a FM NiCo interfacial layer. $Ni_{0.33}Co_{0.67}O$, Gd, and Pd were all fit using a constant value for the real component of the scattering length density, $\rho_N$, and the imaginary component of the scattering length density, $\rho_i$, through the entire thickness of the layers. Even though $\chi^2$ indicates that the overall fit is reasonably good, the spin asymmetry shows that this model does not accurately describe the magnetic contributions from the film.

Figure S2b was fit with a FM NiCo interfacial layer. $Ni_{0.33}Co_{0.67}O$, Gd, and Pd values were kept, as in the previous model, uniform across the entire layer thickness. $\chi^2$ improves slightly with the NiCo layer, and the calculated spin asymmetry is closer to the collected data. This indicates that a FM layer must be included.

Figure S2c shows the fit model, including the FM NiCo interfacial layer, keeping the values constant, as in the previous two models, for $Ni_{0.33}Co_{0.67}O$, Gd, and Pd. The novelty here is the inclusion of a second layer at the interface which possess a higher $\rho_N$ and lower $\rho_i$, corresponding to a more oxidized layer of Gd at the interface. Introducing this layer the model improves significantly (from $\chi^2 = 2.0$ to 1.7). But this model still does not have the best match to the spin asymmetry.

The model shown in Figure S2d was fit again with a FM NiCo interfacial layer. $Ni_{0.34}Co_{0.66}O$ and Pd were all fit using a continuous value for $\rho_N$ and $\rho_i$ for the entire thickness of the layers. In this last example, the difference relies in that a linear gradient $\rho_N$ and $\rho_i$ through 60% of the Gd bulk from the interface was included. The physical meaning of this gradient would correspond to a gradient in oxygen content from the interface into Gd. This model has similar $\chi^2$ values as the third model, which included the $GdO_x$ layer at the interface, but agreement with the spin asymmetry was not further improved. The best model was found by using a layer corresponding to $GdO_x$ above NiCo, similar to the model in Figure S2c, and adding a second magnetic component in Gd near the $Gd/GdO_x$ interface. This second magnetic layer in Gd may be explained as a FM contribution from Gd due to the ambient temperature during measurement being close to the Curie temperature, $T_C$, of Gd (292 K). While not uniform throughout the layer, expansively stained Gd is known to have a surface $T_C$ above the bulk value,[4] which may explain the magnetization existing predominantly at the $Gd/GdO_x$ interface. The fit and spin asymmetry for the accepted AG model are shown in Figure S1a and Fig. 5b, respectively.

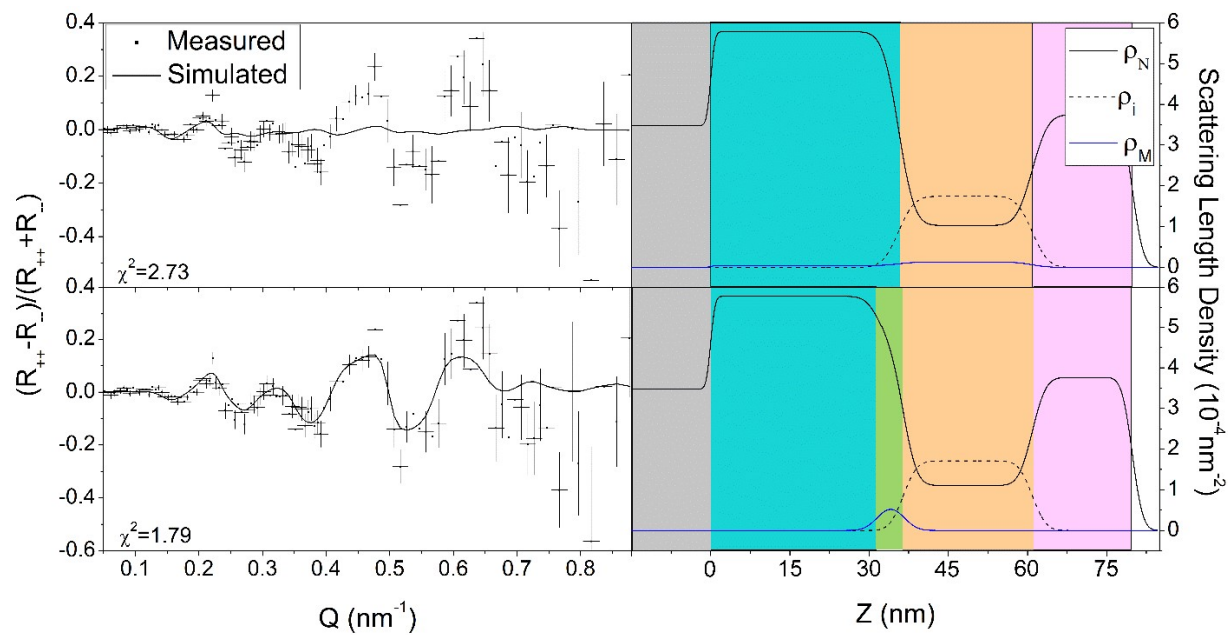

**Figure S3:** Spin asymmetry (left column) and calculated profiles (right column) for different fitting configurations for the $Ni_{0.33}Co_{0.67}O/Gd/Pd$ system in the field-cooled state. (Top) Model not including a FM NiCo interfacial layer, and (bottom) including a NiCo layer. Colored regions are: $SiO_2$ (grey), $Ni_{0.33}Co_{0.67}O$ (teal), NiCo (green), Gd (light orange), Pd (magenta) and air (white). Error bars correspond to ±1 s.d.

Figure S3 shows the spin asymmetry (left column) and calculated profiles (right column) of 2 alternative models that were used during the fitting process of the field cooled, FC, sample. As for the AG sample, the presence of the FM layer must be tested. The top of Figure S3 shows the results for a model not including the FM NiCo interfacial layer, while $Ni_{0.33}Co_{0.67}O$, Gd, and Pd are treated with continuous values of $\rho_N$ and $\rho_i$ throughout the corresponding layers. The poor agreement with the spin asymmetry confirms the necessity to include the FM NiCo layer. The bottom of Figure S3 shows a second model presenting a single FM NiCo layer and the same criteria for $Ni_{0.33}Co_{0.67}O$, Gd and Pd. Compared to the accepted model, this model lacks a GdNi layer which is confirmed in EELS measurement (Fig. 4) and has a higher value of $\chi^2$. Based on the quality of the fit data and additional information obtained using EELS mapping, the accepted model presented in the text must contain a FM layer and a GdNi intermixing layer with the fits and spin asymmetry shown in Figure S1b and Fig. 5d, respectively.

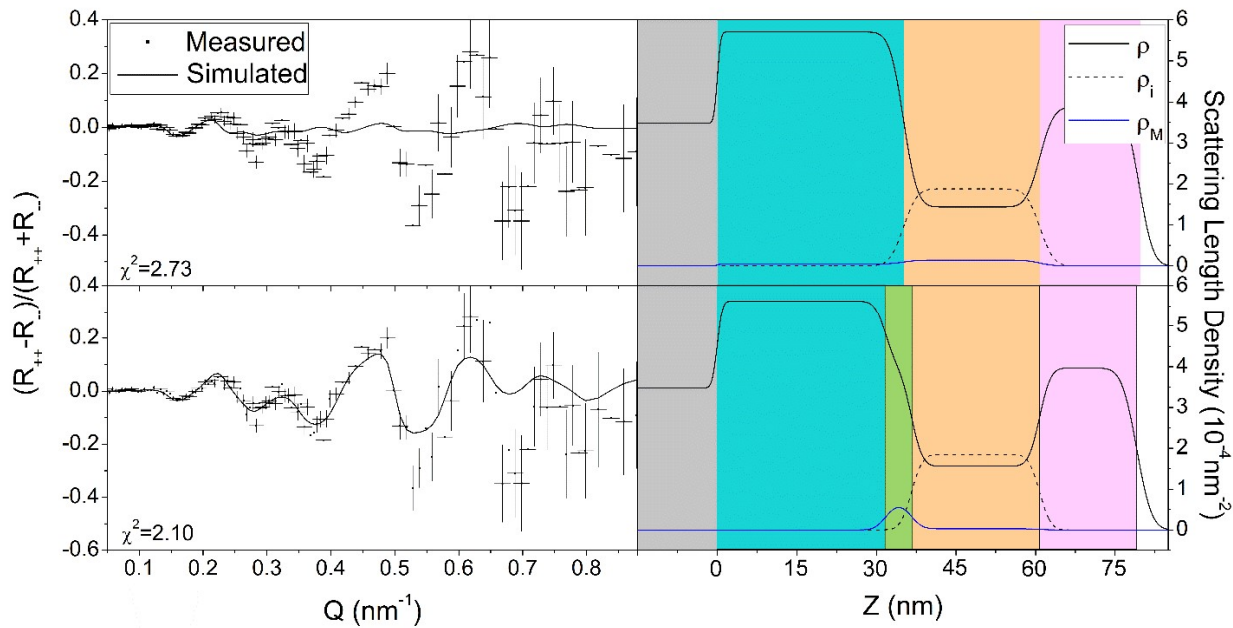

**Figure S4:** Spin asymmetry (left column) and calculated profiles (right column) for different fitting configurations for the $Ni_{0.33}Co_{0.67}O$/Gd/Pd system after the voltage conditioning. In (top) model not including a FM NiCo interfacial layer and in (bottom) with the inclusion of a NiCo layer. Colored regions correspond to: $SiO_2$ (grey), $Ni_{0.33}Co_{0.67}O$ (teal), NiCo reen), Gd (light orange), $GdO_x$ (dark orange), Pd (magenta) and air (white). Error bars correspond to ±1 s.d.

**Supporting Information**

Figure S4 shows the spin asymmetry (left column) and calculated profiles (right column) of 2 alternative models that were used during the fitting process of the field cool and voltage conditioned, FC+VC, sample. The process used to converge in an accepted model was similar to the FC sample. The top of Figure S4 shows the fitting for the Gd/Ni$_{0.33}$Co$_{0.67}$O system without including the NiCo layer, treating with continuous values of $\rho_N$ and $\rho_i$ throughout the corresponding layers. Once again, the lack of spin asymmetry and theory agreement confirms the necessity of its inclusion. The presence of a NiCo interfacial layer was also tested (see bottom of Figure S4). Again, the inclusion of GdNi in the accepted model improves the value of $\chi^2$ while also being confirmed by EELS mapping (Fig. 4). The accepted fit and spin asymmetry are shown in Figure S1c and Fig. 5f, respectively.

**X-Ray Reflectivity**

As shown in Figure S5, XRR of a Ni$_{0.33}$Co$_{0.67}$O (40nm) film reveals many thickness fringes, indicating a good layer structure and a surface roughness of 1.8nm; a comparison sample of as-grown Ni$_{0.33}$Co$_{0.67}$O (40nm)/Gd(20nm)/Pd(20nm), on the other hand, exhibits significantly suppressed reflectivity, consistent with the ion migration-induced disordering.

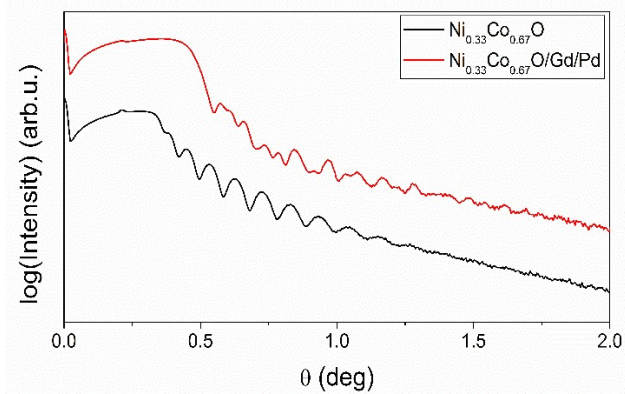

**Figure S5**. X-ray reflectivity measurement of a Ni$_{0.33}$Co$_{0.67}$O (40nm) vs Ni$_{0.33}$Co$_{0.67}$O (40nm)/ Gd(20nm)/ Pd(20nm). The curves are displaced along the intensity axis for clarity.

**Supporting Information**